\magnification \magstep1
\raggedbottom
\openup 2\jot
\voffset6truemm
\def\rvec#1{\vbox{\ialign{##\crcr$\rightarrow$\crcr\noalign{
\kern-1pt \nointerlineskip}$\hfil\displaystyle{#1}\hfil$\crcr}}}
\def\lvec#1{\vbox{\ialign{##\crcr$\leftarrow$\crcr\noalign{
\kern-1pt \nointerlineskip}$\hfil\displaystyle{#1}\hfil$\crcr}}}
\def\bivec#1{\vbox{\ialign{##\crcr$\leftrightarrow$\crcr\noalign{
\kern-1pt \nointerlineskip}$\hfil\displaystyle{#1}\hfil$\crcr}}}
\def\II{{\rm1\!\hskip-1pt I}}

\centerline {\bf PEIERLS BRACKETS IN THEORETICAL PHYSICS}
\vskip 1cm
\noindent
Giampiero Esposito$^{1,2}$, Giuseppe Marmo$^{2,1}$
and Cosimo Stornaiolo$^{1,2}$
\vskip 1cm
\noindent
${ }^{1}${\it INFN, Sezione di Napoli, Complesso Universitario di
Monte S. Angelo, Via Cintia, Edificio N', 80126 Napoli, Italy}
\vskip 0.3cm
\noindent
${ }^{2}${\it Dipartimento di Scienze Fisiche, Universit\`a di
Napoli Federico II, Complesso Universitario di Monte S. Angelo,
Via Cintia, Edificio N', 80126 Napoli, Italy}
\vskip 1cm
\noindent
{\bf Abstract}. Peierls brackets are part of the space-time approach to
quantum field theory, and provide a Poisson bracket which, being
defined for pairs of observables which are group invariant, is
group invariant by construction. It is therefore well suited for
combining the use of Poisson brackets and the full diffeomorphism
group in general relativity. The present paper provides an
introduction to the topic, with applications to field theory
and point Lagrangians.
\vskip 100cm
\leftline {\bf 1. Introduction}
\vskip 0.3cm
\noindent
Although the Hamiltonian formalism provides a powerful tool for
studying general relativity [1], its initial-value problem and
the approach to canonical quantization [2], it suffers from
severe drawbacks: the space $+$ time split of $(M,g)$ disagrees
with the aims of general relativity, and the space-time topology
is taken to be $\Sigma \times {\bf R}$, so that the full diffeomorphism
group of $M$ is lost [3,4].

However, as was shown by DeWitt in the sixties [5], it remains
possible to use a Poisson-bracket formalism which preserves the
full invariance properties of the original theory, by relying upon
the work of Peierls [6]. In our paper, whose aims are pedagogical,
we begin by describing the general framework, assuming that the
reader has just been introduced to the DeWitt condensed notation
[5]. Let us therefore consider, in field theory, disturbances
which satisfy the homogeneous equation
$S_{,ij}\delta \varphi^{j}=0$, $S$ being the classical
action functional. On using
the DeWitt super-condensed notation we write therefore
$$
S_{2}\delta \varphi=0.
\eqno (1.1)
$$
Hereafter $R_{\; \alpha}^{i}$ are the generators of infinitesimal
gauge transformations, with associated
$$
R_{i \alpha} \equiv \gamma_{ij} R_{\; \alpha}^{j}
$$
built from a local and symmetric matrix $\gamma_{ij}$ which
is taken to transform like $S_{,ij}$ under group transformations.
We also consider
$$
R_{i}^{\; \alpha} \equiv {\widetilde \gamma}^{\alpha \beta}
R_{i \beta},
$$
where by hypothesis the matrix ${\widetilde \gamma}^{\alpha \beta}$
is local, non-singular, symmetric, and transforms according to
the adjoint representation of the infinite-dimensional
invariance group.

We expect to impose supplementary conditions on infinitesimal
disturbances, chosen in the form ($R^{t}$ being the transpose of
generators of infinitesimal gauge transformations)
$$
R^{t} \; \gamma \; \delta \varphi=0.
\eqno (1.2)
$$
Thus, on defining the operator
$$
F \equiv S_{2}+\gamma \; R \; {\widetilde \gamma}^{-1}
\; R^{t} \; \gamma ,
\eqno (1.3)
$$
we look for $\delta \varphi$ solving the homogeneous equation
$$
F \delta \varphi=0,
\eqno (1.4)
$$
and we expect to determine $\delta \varphi$ throughout space-time
if it and its derivatives are specified on any spacelike
hypersurface $\Sigma$. Now on defining
$$
{\widetilde G} \equiv G^{+}-G^{-},
\eqno (1.5)
$$
we arrive at an integral formula for $\delta \varphi$, i.e.
$$
\delta \varphi=\int_{\Sigma}{\widetilde G}
{\bivec f}^{\mu}\delta \varphi
d\Sigma_{\mu}.
\eqno (1.6)
$$
The advanced and retarded
Green functions $G^{\pm}$ are left inverses of $-F$:
$$
G^{\pm}{\lvec F}=-\II \Longrightarrow
{\widetilde G}{\lvec F}=0.
\eqno (1.7)
$$
Furthermore, the form of $F$ and arbitrariness of Cauchy data
${\bivec f}^{\mu}\delta \varphi$ imply that $G^{\pm}$ are right inverses
as well, i.e.
$$
{\rvec F}G^{\pm}=-\II \Longrightarrow {\rvec F}{\widetilde G}=0.
\eqno (1.8)
$$
If symmetry of $F$ is required, one also finds
$$
(G^{\pm})^{t}=G^{\mp},
\eqno (1.9a)
$$
and hence
$$
({\widetilde G})^{t}=-{\widetilde G}.
\eqno (1.10a)
$$
When indices are used, the above properties read
$$
G^{+ij}=G^{-ji}, \; G^{-ij}=G^{+ji},
\eqno (1.9b)
$$
and
$$
{\widetilde G}^{ij}=-{\widetilde G}^{ji},
\eqno (1.10b)
$$
because
$$
G^{\pm ij}-G^{\mp ji}=G^{\pm ik}(F_{kl}-F_{lk})G^{\mp jl},
\eqno (1.11)
$$
and $F_{kl}$ is symmetric. These properties show that, on defining
$\delta_{A}^{\pm}B \equiv \varepsilon B_{,i}G^{\pm ij}A_{,j}$,
one has, on relabelling dummy indices,
$$
\delta_{A}^{\pm}B
=\varepsilon B_{,j}G^{\pm ji}A_{,i}
=\varepsilon A_{,i}G^{\mp ij}B_{,j}
=\delta_{B}^{\mp}A.
\eqno (1.12)
$$
These are the {\it reciprocity relations}, which express the idea
that the retarded (resp. advanced) effect of
$A$ on $B$ equals the advanced (resp. retarded) effect of $B$
on $A$. Another cornerstone of the formalism is a relation
involving the Green function $\widehat G$ of the operator
$-{\widehat F}$, having set $R_{k \beta}R_{\; \alpha}^{k}
\equiv {\widehat F}_{\beta \alpha}$; this is
$$
R \; {\widehat G}^{\pm} \; {\widetilde \gamma}
=G^{\pm} \; \gamma \; R,
\eqno (1.13a)
$$
which, on using indices in the condensed notation, reads
$$
R_{\; \alpha}^{i} \; {\widehat G}^{\pm \alpha \beta} \;
{\widetilde \gamma}_{\beta \delta}
=R_{\; \alpha}^{i} \; {\widehat G}_{\; \; \; \delta}^{\pm \alpha}
=G^{\pm ij} \; \gamma_{jk} \; R_{\; \delta}^{k}
=G^{\pm ij} \; R_{j \delta}.
\eqno (1.13b)
$$
This holds because, {\it for background fields satisfying the field
equations}, one finds that
$$
F_{ik} R_{\; \alpha}^{k}=R_{i}^{\; \beta}R_{k \beta}R_{\; \alpha}^{k}
=R_{i}^{\; \beta}{\widehat F}_{\beta \alpha}.
\eqno (1.14)
$$
On multiplying this equation on the left by $G^{\pm ji}$ and on the
right by ${\widehat G}^{\pm \alpha \beta}$ one gets
$$
R_{\; \alpha}^{j}{\widehat G}^{\pm \alpha \beta}
=G^{\pm ji}R_{i}^{\; \beta},
\eqno (1.15)
$$
i.e. the desired formula (1.13b) is proved. Moreover, by virtue of
(1.9b), the transposed equations
$$
{\widehat G}^{\pm \alpha \beta}R_{\; \beta}^{j}
=R_{i}^{\; \alpha}G^{\pm ij}
\eqno (1.16)
$$
also hold.
We are now in a position to define the Peierls bracket of any two
observables $A$ and $B$. First, we consider the operation
$$
D_{A}B \equiv \lim_{\varepsilon \to 0}\varepsilon^{-1}
\delta_{A}^{-}B,
\eqno (1.17)
$$
with $D_{B}A$ obtained by interchanging $A$ with $B$ in (1.17).
The {\it Peierls bracket} of $A$ and $B$ is then defined by
$$
(A,B) \equiv D_{A}B-D_{B}A=\lim_{\varepsilon \to 0}
{1\over \varepsilon}\Bigr[\varepsilon A_{1}G^{+}B_{1}
-\varepsilon A_{1}G^{-}B_{1}\Bigr]
=A_{1}{\widetilde G}B_{1}=A_{,i}{\widetilde G}^{ij}B_{,j},
\eqno (1.18)
$$
where we have used (1.12) and (1.17) to obtain the last expression.
Following DeWitt [7], it should be stressed that the Peierls
bracket depends only on the behaviour of infinitesimal disturbances.
This is not the same, however, as saying that quantum theory is a
theory of infinitesimal disturbances of the underlying classical
theory! This view would not take into account factor ordering
problems in the evaluation of quantum commutators, nor the existence
of non-classical phase effects. Nevertheless, DeWitt could show that
quantum theory can be regarded as a theory of ``finite but small''
disturbances of the classical theory, and he stressed that the exact
theory is indeed completely determined by the behaviour of
infinitesimal disturbances.

In classical mechanics, following Peierls [6], we may arrive at the
derivatives in (1.17) and (1.18) starting from the action functional
$S \equiv \int L \; d\tau$ and considering the extremals of $S$ and
those of $S + \lambda A$, where $\lambda$ is an infinitesimal
parameter and $A$ any function of the path $\gamma$. Next we
consider solutions of the modified equations as expansions in powers
of $\lambda$, and hence the new set of solutions to first order
reads
$$
\gamma'(\tau)=\gamma(\tau)+\lambda D_{A} \gamma(\tau).
\eqno (1.19)
$$
This modified solution is required to obey the condition
that, in the distant past, it should be identical with the original
one, i.e.
$$
D_{A}\gamma(\tau) \rightarrow 0 \; {\rm as} \; \tau
\rightarrow -\infty.
\eqno (1.20)
$$
Similarly to the construction of the above ``retarded'' solution,
we may define an ``advanced'' solution
$$
\gamma''(\tau)=\gamma(\tau)+\lambda {\cal D}_{A} \gamma(\tau),
\eqno (1.21)
$$
such that
$$
{\cal D}_{A}\gamma(\tau) \rightarrow 0 \; {\rm as} \;
\tau \rightarrow + \infty.
\eqno (1.22)
$$
{}From these modified solutions one can now find $D_{A}\gamma(\tau)$
along the solutions of the un-modified action and therefore,
to first order,
the changes in any other function $B$ of the field variables, and
these are denoted by $D_{A}B$ and $D_{B}A$, respectively.
\vskip 0.3cm
\leftline {\bf 2. Mathematical properties of Peierls brackets}
\vskip 0.3cm
\noindent
We are now aiming to prove that
$(A,B)$ satisfies all properties of a Poisson bracket. The first
two are indeed obvious:
$$
(A,B)=-(B,A),
\eqno (2.1)
$$
$$
(A,B+C)=(A,B)+(A,C),
\eqno (2.2)
$$
whereas the proof of the Jacobi identity is not obvious and is
therefore presented in detail. First, by repeated application of
(1.18) one finds
$$ \eqalignno{
\; & P(A,B,C) \equiv (A,(B,C))+(B,(C,A))+(C,(A,B)) \cr
&=A_{,i}{\widetilde G}^{il}\Bigr(B_{,j}{\widetilde G}^{jk}
C_{,k}\Bigr)_{,l}+B_{,j}{\widetilde G}^{jl}
\Bigr(C_{,k}{\widetilde G}^{ki}A_{,i}\Bigr)_{,l}
+C_{,k}{\widetilde G}^{kl}
\Bigr(A_{,i}{\widetilde G}^{ij}B_{,j}\Bigr)_{,l} \cr
&=A_{,il}B_{,j}C_{,k}\Bigr({\widetilde G}^{ij}
{\widetilde G}^{kl}+{\widetilde G}^{jl}{\widetilde G}^{ki}\Bigr)
+A_{,i}B_{,jl}C_{,k}\Bigr({\widetilde G}^{jk}
{\widetilde G}^{il}+{\widetilde G}^{kl}{\widetilde G}^{ij}\Bigr) \cr
&+A_{,i}B_{,j}C_{,kl}\Bigr({\widetilde G}^{ki}
{\widetilde G}^{jl}+{\widetilde G}^{il}{\widetilde G}^{jk}\Bigr) \cr
&+A_{,i}B_{,j}C_{,k}\Bigr({\widetilde G}^{il}
{\widetilde G}_{\; \; \; ,l}^{jk}
+{\widetilde G}^{jl}{\widetilde G}_{\; \; \; ,l}^{ki}
+{\widetilde G}^{kl}{\widetilde G}_{\; \; \; ,l}^{ij}\Bigr).
&(2.3)\cr}
$$
Now the antisymmetry property (1.10b), jointly with commutation
of functional derivatives: $T_{,il}=T_{,li}$ for all $T=A,B,C$,
implies that the first three terms on the last equality in (2.3)
vanish. For example one finds
$$ \eqalignno{
\; & A_{,il}B_{,j}C_{,k}\Bigr({\widetilde G}^{ij}
{\widetilde G}^{kl}+{\widetilde G}^{jl}{\widetilde G}^{ki}\Bigr)
=A_{,li}B_{,j}C_{,k}\Bigr({\widetilde G}^{lj}
{\widetilde G}^{ki}+{\widetilde G}^{ji}{\widetilde G}^{kl}\Bigr)\cr
&=-A_{,il}B_{,j}C_{,k}\Bigr({\widetilde G}^{jl}
{\widetilde G}^{ki}+{\widetilde G}^{ij}{\widetilde G}^{kl}\Bigr)=0,
&(2.4)\cr}
$$
and an entirely analogous procedure can be applied to the terms
containing the second functional derivatives $B_{,jl}$ and
$C_{,kl}$. The last term in (2.3) requires new calculations
because it contains functional derivatives of ${\widetilde G}^{ij}$.
These can be dealt with after taking infinitesimal variations of
the equation $F_{ik}G^{\pm kj}=-\delta_{i}^{\; j}$, so that
$$
F \; \delta G^{\pm}=-(\delta F)G^{\pm},
\eqno (2.5)
$$
and hence
$$
G^{\pm}F \delta G^{\pm}=F G^{\pm} \delta G^{\pm}
=-\delta G^{\pm}=-G^{\pm}(\delta F)G^{\pm},
\eqno (2.6a)
$$
i.e.
$$
\delta G^{\pm}=G^{\pm}(\delta F)G^{\pm}.
\eqno (2.6b)
$$
Thus, the desired functional derivatives of advanced and retarded
Green functions read
$$ \eqalignno{
\; & G_{\; \; \; \; \; ,c}^{\pm ij}=G^{\pm ia}F_{ab,c}G^{\pm bj}
=G^{\pm ia}\Bigr(S_{,ab}+R_{a \alpha}R_{b}^{\; \alpha}\Bigr)_{,c}
G^{\pm bj} \cr
&=G^{\pm ia}S_{,abc}G^{\pm bj}+G^{\pm ia}R_{a \alpha,c}
R_{b}^{\; \alpha} \; G^{\pm bj}
+G^{\pm ia}R_{a \alpha} R_{b \; \; ,c}^{\; \alpha}
\; G^{\pm bj}.
&(2.7)\cr}
$$
In this formula the contractions $R_{b}^{\; \alpha} \; G^{\pm bj}$
and $G^{\pm ia}R_{a \alpha}$ can be re-expressed with the help of
Eqs. (1.15) and (1.16), and eventually one gets
$$
G_{\; \; \; \; \; ,c}^{\pm ij}=G^{\pm ia} S_{,abc}G^{\pm bj}
+G^{\pm ia}R_{a \alpha,c}{\widehat G}^{\pm \alpha \beta}
R_{\; \beta}^{j}
+R_{\; \beta}^{i} \; {\widehat G}_{\; \; \; \alpha}^{\pm \beta}
\; R_{b \; \; ,c}^{\; \alpha} \; G^{\pm bj}.
\eqno (2.8)
$$
By virtue of the group invariance property satisfied by all
physical observables, the second and third term on the right-hand
side of Eq. (2.8) give vanishing contribution to (2.3). One is
therefore left with the contributions involving third functional
derivatives of the action. Bearing in mind that $S_{,abc}=
S_{,acb}=S_{,bca}=...$, one can relabel indices summed over, finding
eventually (upon using (1.9b))
$$ \eqalignno{
\; & P(A,B,C)=A_{,i}B_{,j}C_{,k}\Bigr[(G^{+ic}-G^{-ic})
(G^{+ja}G^{+bk}-G^{-ja}G^{-bk}) \cr
&+(G^{+jc}-G^{-jc})(G^{+ka}G^{+bi}-G^{-ka}G^{-bi}) \cr
&+(G^{+kc}-G^{-kc})(G^{+ia}G^{+bj}-G^{-ia}G^{-bj})\Bigr]S_{,abc} \cr
&=A_{,i}B_{,j}C_{,k}\Bigr[(G^{+ia}-G^{-ia})(G^{+jb}G^{-kc}
-G^{-jb}G^{+kc})\cr
&+(G^{+jb}-G^{-jb})(G^{+kc}G^{-ia}-G^{-kc}G^{+ia})\cr
&+(G^{+kc}-G^{-kc})(G^{+ia}G^{-jb}-G^{-ia}G^{+jb})
\Bigr]S_{,abc} =0.
&(2.9)\cr}
$$
This sum vanishes because it involves six pairs of triple products of
Green functions with opposite signs, i.e.
$$
G^{+ia}G^{+jb}G^{-kc}, \; G^{-ia}G^{-jb}G^{+kc}, \;
G^{+jb}G^{+kc}G^{-ia},
$$
$$
G^{-jb}G^{-kc}G^{+ia}, \; G^{+kc}G^{+ia}G^{-jb}, \;
G^{-kc}G^{-ia}G^{+jb}.
$$
The Jacobi identity is therefore fulfilled. Moreover, the fourth
fundamental property of Poisson brackets, i.e.
$$
(A,BC)=(A,B)C+B(A,C)
\eqno (2.10)
$$
is also satisfied, because
$$
(A,BC)=A_{,i}{\widetilde G}^{ik}(BC)_{,k}
=A_{,i}{\widetilde G}^{ik}B_{,k}C+BA_{,i}{\widetilde G}^{ik}C_{,k}
=(A,B)C+B(A,C).
\eqno (2.11)
$$
Thus, the Peierls bracket defined in (1.18) is indeed a Poisson
bracket of physical observables. Equation (2.10) can be regarded
as a compatibility condition of the Peierls bracket with the product
of physical observables.

It should be stressed that the idea of Peierls [6] was to introduce
a bracket related directly to the action principle without making
any reference to the Hamiltonian. This implies that even classical
mechanics should be considered as a ``field theory'' in a
zero-dimensional space, having only the time dimension. This means
that one deals with an infinite-dimensional space of paths
$\gamma: {\bf R} \rightarrow Q$, therefore we are dealing with
functional derivatives and distributions even in this situation
where modern standard treatments rely upon $C^{\infty}$ manifolds
and smooth structures. Thus, the present treatment is hiding most
technicalities involving infinite-dimensional manifolds. In finite
dimensions on a smooth manifold, any bracket satisfying (2.2) and
(2.10) is associated with first-order bidifferential operators [8,9];
in this proof it is important that the commutative and associative
product $BC$ is a local product. In any case these brackets at the
classical level could be a starting point to define a
$*$-product in the spirit of non-commutative geometry [10] or
deformation quantization [11].
\vskip 5cm
\leftline {\bf 2.1 The most general Peierls bracket}
\vskip 0.3cm
\noindent
The Peierls bracket is a group invariant by construction, being
defined for pairs of observables which are group invariant, and
is invariant under both infinitesimal and finite changes in the
matrices $\gamma_{ij}$ and ${\widetilde \gamma}_{\alpha \beta}$.
DeWitt [5] went on to prove that, even if
independent differential operators $P_{i}^{\; \alpha}$ and
$Q_{i \alpha}$ are introduced such that
$$
F_{ij} \equiv S_{,ij}+P_{i}^{\; \alpha}Q_{j \alpha},
\eqno (2.12)
$$
$$
{\widehat F}_{\alpha \beta} \equiv Q_{i \alpha}R_{\; \beta}^{i},
\eqno (2.13)
$$
$$
F_{\alpha}^{\; \beta} \equiv R_{\; \alpha}^{i}
P_{i}^{\; \beta},
\eqno (2.14)
$$
are all non-singular, with unique advanced and retarded Green
functions, the reciprocity theorem expressed by (1.12) still
holds, and the resulting Peierls bracket is invariant under changes
in the $P_{i}^{\; \alpha}$ and $Q_{i \alpha}$, by virtue of
the identities
$$
Q_{i \alpha}G^{\pm ij}=G_{\; \; \alpha}^{\pm \; \; \beta}
R_{\; \beta}^{j},
\eqno (2.15)
$$
$$
G^{\pm ij}P_{j}^{\; \beta}=R_{\; \alpha}^{i}
{\widehat G}^{\pm \alpha \beta}.
\eqno (2.16)
$$
This is proved as follows. The composition of $F_{ik}$ with the
infinitesimal generators of gauge transformations yields
$$
F_{ik}R_{\; \alpha}^{k}=P_{i}^{\; \beta}F_{\beta \alpha},
\eqno (2.17)
$$
and hence
$$
G^{\pm ji}F_{ik}R_{\; \alpha}^{k}=-R_{\; \alpha}^{j}
=G^{\pm ji}P_{i}^{\; \gamma}F_{\gamma \alpha},
\eqno (2.18)
$$
which implies
$$
R_{\; \alpha}^{j}G^{\pm \alpha \beta}=-G^{\pm ji}P_{i}^{\; \gamma}
F_{\gamma \alpha}G^{\pm \alpha \beta}
=G^{\pm ji}P_{i}^{\; \beta},
\eqno (2.19)
$$
i.e. Eq. (2.16) is obtained. Similarly,
$$
R_{\; \alpha}^{i}F_{ij}=F_{\alpha}^{\; \beta}Q_{j \beta},
\eqno (2.20)
$$
and hence
$$
G_{\; \alpha}^{\pm \; \; \gamma}R_{\; \gamma}^{i}F_{ij}
=-Q_{j \alpha},
\eqno (2.21)
$$
which implies
$$
Q_{i \alpha}G^{\pm ij}=-G_{\; \alpha}^{\pm \; \; \gamma}
R_{\; \gamma}^{k}F_{ki}G^{\pm ij}
=G_{\alpha}^{\pm \; \; \beta}R_{\; \beta}^{j},
\eqno (2.22)
$$
i.e. Eq. (2.15) is obtained. Now we use the first line of Eq. (1.12)
for $\delta_{A}^{\pm}B$, jointly with
$$
G^{\pm ij}=G^{\mp ji}+G^{\pm ik}(F_{kl}-F_{lk})G^{\mp jl},
\eqno (2.23)
$$
so that
$$
\delta_{A}^{\pm}B-\varepsilon B_{,i}G^{\mp ji}A_{,j}
=\varepsilon B_{,i}R_{\; \gamma}^{i}G^{\pm \gamma \alpha}
Q_{l \alpha}G^{\mp jl}A_{,j}
-\varepsilon B_{,i}P_{l}^{\; \alpha}G^{\pm ik}Q_{k \alpha}
G^{\mp jl}A_{,j}.
\eqno (2.24)
$$
Since $B$ is an observable by hypothesis, the first term on the right-hand
side of (2.24) vanishes. Moreover one finds, from (2.16)
$$
G^{\pm ik}P_{l}^{\; \alpha}Q_{k \alpha}G^{\mp jl}
=G^{\pm il}R_{\; \beta}^{j}G^{\mp \beta \alpha}Q_{l \alpha}.
\eqno (2.25)
$$
and hence also the second term on the right-hand side of (2.24) vanishes
($A$ being an observable, for which $R_{\; \beta}^{j}A_{,j}=0$), yielding
eventually the reciprocity relation (1.12). Moreover, the invariance of the
Peierls bracket under variations of $P_{i \alpha}$ and $Q_{i}^{\; \alpha}$
holds because
$$ \eqalignno{
\; & \delta (\delta_{A}^{\pm}B)=\varepsilon B_{,i}\delta G^{\pm ij}A_{,j}
=\varepsilon B_{,i}G^{\pm ik}(\delta F_{kl})G^{\pm lj}A_{,j} \cr
&=\varepsilon B_{,i}G^{\pm ik}\Bigr[(\delta P_{k}^{\; \alpha})Q_{l \alpha}
+P_{k}^{\; \alpha}(\delta Q_{l \alpha})\Bigr]G^{\pm lj}A_{,j} \cr
&=\varepsilon B_{,i}G^{\pm ik}(\delta P_{k}^{\; \alpha})Q_{l \alpha}
G^{\pm lj}A_{,j}+\varepsilon B_{,i}G^{\pm ik}P_{k}^{\; \alpha}
(\delta Q_{l \alpha})G^{\pm lj}A_{,j} \cr
&=\varepsilon B_{,i}G^{\pm ik} (\delta P_{k}^{\; \alpha})
G_{\; \alpha}^{\pm \; \; \beta}R_{\; \beta}^{j}A_{,j}
+\varepsilon B_{,i}R_{\; \gamma}^{i}G^{\pm \gamma \alpha}
(\delta Q_{l \alpha})G^{\pm lj}A_{,j}=0,
&(2.26)\cr}
$$
where Eqs. (2.15) and (2.16) have been exploited once more.
\vskip 1cm
\leftline {\bf 3. Point Lagrangians}
\vskip 0.3cm
\noindent
A basic question is whether the Peierls-bracket formalism is
equivalent to the conventional canonical formalism when the
latter exists. This is indeed the case, and the proof is given
as follows in the case of point Lagrangians, relying upon the
work in Ref. [5].

Let us consider a physical system possessing only a finite number
of degrees of freedom, with Lagrangian $L$ depending on
positions $q$ and velocities $v$.
We use second-order formalism, {\it assuming} therefore that
$$
v^{i}={d\over dt}q^{i}={\dot q}^{i}.
\eqno (3.1)
$$
We also assume that the equations defining canonical momenta
$p_{i}$ as derivatives of $L$ with respect to ${\dot q}^{i}$
can be solved for ${\dot q}^{i}$ in terms of the $p_{i}$ and
$q^{i}$, so that the Hessian matrix is non-singular. Our
action functional is therefore
$$
S=\int L(q,\dot{q})dt,
$$
whose second variation reads
$$ \eqalignno{
\; & \delta^{2}S=\delta (\delta S)=\delta \int \left[  {{\delta
 L}\over{\delta q^{i}}}\delta q^{i} +  {{\delta L}\over{\delta \dot{q}
^{i}}}\delta \dot{q}^{i}\right] \cr
&=\int\left[{{\delta^{2}L}\over{\delta q^{i}\delta q^{j'}}}\delta
q^{i}\delta q^{j'} +{{\delta^{2}L}\over{\delta q^{i}\delta
\dot{q}^{j'}} }\delta q^{i}\delta
\dot{q}^{j'}+{{\delta^{2}L}\over{\delta \dot{q}^{i}\delta q^{j'}}
}\delta\dot{q}^{i}\delta q^{j'}+{{\delta^{2}L}\over{\delta
\dot{q}^{i}\delta \dot{q}^{j'}} }\delta \dot{q}^{i}\delta
\dot{q}^{j'}\right]
&(3.2)\cr}
$$
If we perform an integration by parts with respect to $\delta
\dot{q}^{i}$, we obtain
$$ \eqalignno{
\; & {\rm integrand\, of\, (3.2)}={{\delta^{2}L}\over{\delta
q^{i}\delta q^{j'}}}\delta q^{i}\delta q^{j'} +
{{\delta^{2}L}\over{\delta q^{i}\delta \dot{q}^{j'}}}\delta
q^{i}\delta {\dot q}^{j'} + {d\over{dt}}\left[{{\delta^{2}L}\over{\delta
\dot{q}^{i}\delta q^{j'}}}\delta q^{i}\delta  q^{j'}\right] \cr
&-{d\over{dt}}\left( {{\delta^{2}L}\over{\delta \dot{q}^{i}\delta
q^{j'}}}\right)\delta q^{i}\delta q^{j'} -
{{\delta^{2}L}\over{\delta \dot{q}^{i}\delta q^{j'}}}\delta
q^{i}\delta \dot{q}^{j'} +
{d\over{dt}}\left[{{\delta^{2}L}\over{\delta \dot{q}^{i}\delta
\dot{q}^{j'}}}\delta q^{i}\delta \dot{q}^{j'}\right] \cr
&-{d\over{dt}}{{\delta^{2}L}\over{\delta \dot{q}^{i}\delta
\dot{q}^{j'}}}\delta q^{i}\delta \dot{q}^{j'}
-{{\delta^{2}L}\over{\delta \dot{q}^{i}\delta
\dot{q}^{j'}}}\delta q^{i}\delta \ddot{q}^{j'} \cr
&=\left[{{\delta^{2}L}\over{\delta q^{i}\delta
q^{j'}}}-{d\over{dt}}{{\delta^{2}L}\over{\delta \dot{q}^{i}\delta
q^{j'}}} \right]\delta q^{i}\delta q^{j'} \cr
&+\left[{{\delta^{2}L}\over{\delta q^{i}\delta
\dot{q}^{j'}}}-{{\delta^{2}L}\over{\delta \dot{q}^{i}\delta
q^{j'}}} -{d\over{dt}}{{\delta^{2}L}\over{\delta \dot{q}^{i}\delta
\dot{q}^{j'}}}\right]\delta q^{i}\delta \dot{q}^{j'} -
{{\delta^{2}L}\over{\delta \dot{q}^{i}\delta \dot{q}^{j'}}}\delta
q^{i}\delta \ddot{q}^{j'} \cr
&+ {d\over{dt}}\left[ {{\delta^{2}L}\over{\delta \dot{q}^{i}\delta
q^{j'}}}\delta q^{i}\delta q^{j'} + {{\delta^{2}L}\over{\delta
\dot{q}^{i}\delta \dot{q}^{j'}}}\delta q^{i}\delta \dot{q}^{j'}
\right].
&(3.3)\cr}
$$
The total derivative can be discarded in the previous equation,
and hence we find
$$
{\rm integrand\, of\, (3.2)}= A_{ij'}\delta q^{i}\delta q^{j'}
+ B_{ij'}\delta q^{i}\delta \dot{q}^{j'} + C_{ij'}\delta q^{i}\delta
\ddot{q}^{j'},
\eqno (3.4)
$$
with obvious definition of $A_{ij'}, B_{ij'}$ and $C_{ij'}$ from
the last three lines of (3.3). We can modify Eq. (3.4)
using only un-primed indices by virtue of the following relations:
$$
\delta q^{j'}={{\partial  q^{j'}(t')}\over{\partial  q^{j}(t)}}
\delta q^{j}= \delta^{j'}_{j}\delta(t,t')\delta q^{j},
\eqno (3.5)
$$
$$
\delta \dot{q}^{j'}=\delta^{j'}_{j}\left[ {\partial
\delta(t,t')\over{\partial t}}\delta q^{j}+\delta(t,t')\delta
\dot{q}^{j}\right],
\eqno (3.6)
$$
$$
\delta \ddot{q}^{j'}=\delta^{j'}_{j}\left[ {\partial^{2}
\delta(t,t')\over{\partial t^{2}}}\delta q^{j}+    2{\partial
\delta(t,t')\over{\partial t}}\delta \dot{q}^{j}+
\delta(t,t')\delta \ddot{q}^{j}\right].
\eqno (3.7)
$$
After substituting into Eq. (3.4), we find eventually
$$ \eqalignno{
\; & {\rm integrand\, of\, (3.2)}=\left[ \left(A_{ij}\delta(t,t')
+ B_{ij}{\partial \delta(t,t')\over{\partial t}}+
C_{ij}{\partial^{2} \delta(t,t')\over{\partial t^{2}}}\right)
\delta q^{j} \right . \cr
& \left . +\left(B_{ij}\delta(t,t') +2C_{ij}{\partial
\delta(t,t')\over \partial t}\right)
\delta {\dot q}^{j}
+ C_{ij}\delta(t,t')\delta \ddot{q}^{j}\right]\delta
q^{i}.
&(3.8)\cr}
$$
The result (3.8) yields the desired formula for the second
functional derivative of the action, i.e.
$$
S_{,ij'}=A_{ij}\delta(t,t')+B_{ij}{\partial \over \partial t}
\delta(t,t')+C_{ij}{\partial^{2}\over \partial t^{2}}
\delta(t,t').
\eqno (3.9)
$$
Since $S_{,ik}G^{\pm kj'}=-\delta_{i}^{\; j'}$, the equation for
the Green function reads
$$
A_{ik}G^{\pm kj'}+B_{ik}{\dot G}^{\pm kj'}
+C_{ik}{\ddot G}^{kj'}=-\delta_{i}^{\; j'}.
\eqno (3.10)
$$
Now it is crucial that $C_{ij} \equiv -{\partial^{2}L \over
\partial {\dot q}^{i} \partial {\dot q}^{j}}$ should have an inverse
according to our assumptions (when this is not the case we are
forced to use the constraint approach [12,13] of Dirac and Bergmann).
This makes it possible to write the
solution of Eq. (3.10) as $|t-t'| \rightarrow 0$ in the form
$$
G^{+ij'}=-\theta(t',t)(t'-t)C^{-1i'j'}+{\rm O}(t-t')^{2},
\eqno (3.11)
$$
$$
G^{-ij'}=-\theta(t,t')(t-t')C^{-1i'j'}+{\rm O}(t-t')^{2},
\eqno (3.12)
$$
so that the ``super-commutator function'' (1.5) is given by
$$
{\widetilde G}^{ij}=(t-t')C^{-1i'j'}+{\rm O}(t-t')^{2}.
\eqno (3.13)
$$
By construction, ${\widetilde G}^{ij}$ solves the homogeneous
equation
$$
A_{ik}{\widetilde G}^{kj'}+B_{ik}{\dot {\widetilde G}}^{kj'}
+C_{ik}{\ddot {\widetilde G}}^{kj'}=0,
\eqno (3.14)
$$
and by inserting (3.13) into (3.14) one finds
$$
B_{i'k'}C^{-1k'j'}+C_{i'k'}{\partial^{2}\over \partial t^{2}}
{\widetilde G}^{k'j'}={\rm O}(t-t'),
\eqno (3.15)
$$
which is solved by (cf. (3.13))
$$
{\widetilde G}^{ij'}=(t-t')C^{-1i'j'}-{1\over 2}(t-t')^{2}
C^{-1i'k'}B_{k'l'}C^{-1l'j'}+{\rm O}(t-t')^{3}.
\eqno (3.16)
$$
Now we need the Peierls brackets $(q^{i},q^{j}), (q^{i},p_{j})$ and
$(p_{i},p_{j})$. The first task is easy, because
$$ \eqalignno{
\; & (q^{i}(t),q^{j}(t))=\lim_{t' \to t}(q^{i}(t),q^{j}(t')) \cr
&=\lim_{t' \to t}{\partial q^{i}\over \partial q^{k}(t)}
{\widetilde G}^{kl'}{\partial q^{j}\over \partial q^{l}(t')}
=\lim_{t' \to t}\delta_{\; k}^{i} {\widetilde G}^{kl'}
\delta_{\; l}^{j'} \cr
&=\lim_{t' \to t}{\widetilde G}^{ij'}=0,
&(3.17)\cr}
$$
by virtue of (3.16). The remaining brackets require an intermediate
step, i.e. evaluation of the Peierls brackets $(q^{i},{\dot q}^{j})$
and $({\dot q}^{i},{\dot q}^{j})$, since for example
$$ \eqalignno{
\; & (p_{i},p_{j})=\left({\partial L \over \partial {\dot q}^{i}},
{\partial L \over \partial {\dot q}^{j}}\right) \cr
&={\partial^{2}L\over \partial {\dot q}^{i} \partial q^{k}}
(q^{k},q^{l}){\partial^{2}L \over \partial q^{l} \partial {\dot q}^{j}}
+{\partial^{2}L \over \partial {\dot q}^{i} \partial q^{k}}
(q^{k},{\dot q}^{l}){\partial^{2}L \over \partial {\dot q}^{l}
\partial {\dot q}^{j}} \cr
&+{\partial^{2}L \over \partial {\dot q}^{i} \partial {\dot q}^{k}}
({\dot q}^{k},q^{l}){\partial^{2}L \over \partial q^{l}
\partial {\dot q}^{j}}+{\partial^{2}L \over \partial {\dot q}^{i}
\partial {\dot q}^{k}}({\dot q}^{k},{\dot q}^{l})
{\partial^{2}L \over \partial {\dot q}^{l} \partial {\dot q}^{j}}.
&(3.18)\cr}
$$
Indeed one finds
$$
(q^{i},{\dot q}^{j})=-({\dot q}^{j},q^{i})
=-\lim_{t' \to t}{\dot {\widetilde G}}^{ji'}=-C^{-1ji}
=-C^{-1ij},
\eqno (3.19)
$$
$$
({\dot q}^{i},{\dot q}^{j})=\lim_{t' \to t}
{\partial^{2}{\widetilde G}^{ij'}\over \partial t \partial t'}
=C^{-1ik}\left(B_{kl}-{dC_{kl}\over dt}\right)C^{-1lj},
\eqno (3.20)
$$
and hence
$$
(q^{i},p_{j})=C^{-1ik}C_{kj}=\delta_{\; j}^{i},
\eqno (3.21)
$$
$$
(p_{i},p_{j})={\partial^{2}L \over \partial {\dot q}^{i}
\partial q^{j}}-{\partial^{2}L \over \partial q^{i}
\partial {\dot q}^{j}}+B_{ij}-{dC_{ij}\over dt}=0.
\eqno (3.22)
$$
Thus, we fully recover the canonical commutation relations (3.17),
(3.21) and (3.22), with Lagrangian sub-spaces corresponding to (3.17)
and (3.22). The familiar Hamilton equations can also be recovered [5].
\vskip 0.3cm
\leftline {\bf 4. Concluding remarks}
\vskip 0.3cm
\noindent
In agreement with the pedagogical aims of the Londrina school, we
have presented a concise introduction to Peierls brackets in
theoretical physics. For this purpose, we find it useful to
supplement the previous discussion with the following correspondence
of structures:
\vskip 0.3cm
\noindent
(i) Finite-dimensional manifold $M$ in classical mechanics vs.
infinite-dimensional manifold $\Phi$ of field configurations.
\vskip 0.3cm
\noindent
(ii) Local coordinates $\left \{ \xi^{i} \right \}$ on $M$ vs. field
configurations $\left \{ \varphi^{i} \right \}$ on $\Phi$.
\vskip 0.3cm
\noindent
(iii) Poisson bracket $\left \{ \xi^{i},\xi^{j} \right \}
=\omega^{ij}$, with $\omega^{ij}$ invertible, vs. the Peierls bracket
$$
(\varphi^{i},\varphi^{j})=\varphi_{\; ,k}^{i}
{\widetilde G}^{kl}\varphi_{\; ,l}^{j}
=\delta_{\; k}^{i}{\widetilde G}^{kl}\delta_{\; l}^{j}
={\widetilde G}^{ij}.
\eqno (4.1)
$$
(iv) Inverse matrix $\omega_{ij}$ such that $\omega_{ij}\omega^{jk}
=\delta_{i}^{\; k}$, with associated symplectic form
$\omega \equiv {1\over 2}\omega_{ij}d\xi^{i} \wedge d\xi^{j}$,
vs. inverse of ${\widetilde G}^{ij}$ built as
$$
{\widetilde G}^{ik}\gamma_{im}\gamma_{kl}
={\widetilde G}_{ml},
\eqno (4.2)
$$
for which ${\widetilde G}_{ml}{\widetilde G}^{lk}=\delta_{m}^{\; \; k}$.
\vskip 0.3cm
\noindent
(v) Symplectic manifold $(M,\omega)$ vs. $(\Phi,{\widetilde G})$.
\vskip 0.3cm
\noindent
(vi) Functions $f$ on $M$, i.e. $f \in {\cal F}(M)$, vs. observables
$A(\varphi)$ on the set ${\cal O}(\Phi)$ of all observables on $\Phi$.
\vskip 0.3cm
\noindent
(vii) Poisson bracket in local coordinates:
$$
\left \{ f,h \right \}={\partial f \over \partial \xi^{i}}
\omega^{ij}{\partial h \over \partial \xi^{j}}
=f_{,i}\omega^{ij}h_{,j}
$$
vs. the Peierls bracket
$$
(A,B)=A_{,i}{\widetilde G}^{ij}B_{,j}=\int dx \int dy
{\delta A \over \delta \varphi^{i}(x)}{\widetilde G}^{ij}(x,y)
{\delta B \over \delta \varphi^{j}(y)}.
\eqno (4.3)
$$
(viii) Differential of $f$, i.e.
$df={\partial f \over \partial \xi^{i}}d\xi^{i}=f_{,i}d\xi^{i}$, vs.
variation of the functional $A(\varphi)$:
$$
\delta A =\int {\delta A \over \delta \varphi^{i}}\delta \varphi^{i}dx
=A_{,i}\delta \varphi^{i}.
\eqno (4.4)
$$
(ix) Poisson bracket $\left \{ \; , \; \right \}:
{\cal F}(M) \times {\cal F}(M) \rightarrow {\cal F}(M)$ vs.
Peierls bracket
$$
(\; , \;): {\cal O}(\Phi) \times {\cal O}(\Phi)
\rightarrow {\cal O}(\Phi).
$$

Current applications of Peierls brackets deal with string theory
[14,15], path integration and decoherence [16], supersymmetric
proof of the index theorem [17], classical dynamical systems
involving parafermionic and parabosonic dynamical variables [18],
while for recent literature on covariant approaches to a
canonical formulation of field theories we refer the reader to
the work in Refs. [19-24].

In the infinite-dimensional setting which, strictly, applies also
to classical mechanics, as we stressed at the end of section 2,
we hope to elucidate the relation between a covariant description
of dynamics as obtained from the kernel of the symplectic form,
and a parametrized description of dynamics as obtained from any
Poisson bracket, including the Peierls bracket.
\vskip 0.3cm
\leftline {\bf Acknowledgments}
\vskip 0.3cm
\noindent
The work of the authors has been partially supported
by PRIN 2000 {\it Sintesi}.
\vskip 0.3cm
\leftline {\bf References}
\vskip 0.3cm
\item {[1]}
P. A. M. Dirac, Phys. Rev. {\bf 114}, 924 (1959).
\item {[2]}
B. S. DeWitt, Phys. Rev. {\bf 160}, 1113 (1967).
\item {[3]}
C. J. Isham and K. Kuchar, Ann. Phys. {\bf 164}, 288 (1985).
\item {[4]}
C. J. Isham and K. Kuchar, Ann. Phys. {\bf 164}, 316 (1985).
\item {[5]}
B. S. DeWitt, {\it Dynamical Theory of Groups and Fields}
(Gordon \& Breach, New York, 1965).
\item {[6]}
R. E. Peierls, Proc. R. Soc. Lond. {\bf A 214}, 143 (1952).
\item {[7]}
B. S. DeWitt, Phys. Rev. {\bf 162}, 1195 (1967).
\item {[8]}
J. Grabowski and G. Marmo, J. Phys. {\bf A 34}, 3803 (2001).
\item {[9]}
J. Grabowski and G. Marmo, ``Binary Operations in
Classical and Quantum Mechanics'' (MATH-DG 0201089).
\item {[10]}
A. Connes, J. Math. Phys. {\bf 41}, 3832 (2000).
\item {[11]}
J. M. Gracia-Bondia, F. Lizzi, G. Marmo and P. Vitale,
JHEP {\bf 0204}, 026 (2002).
\item {[12]}
G. Marmo, N. Mukunda and J. Samuel, Riv. Nuovo Cimento
{\bf 6}, 1 (1983).
\item {[13]}
B. A. Dubrovin, M. Giordano, G. Marmo and A. Simoni,
Int. J. Mod. Phys. {\bf A 8}, 3747 (1993).
\item {[14]}
S. R. Das, C. R. Ordonez and M. A. Rubin, Phys. Lett.
{\bf B 195}, 139 (1987).
\item {[15]}
C. R. Ordonez and M. A. Rubin, Phys. Lett. {\bf B 216},
117 (1989).
\item {[16]}
D. M. Marolf, Ann. Phys. (N.Y.) {\bf 236}, 392 (1994).
\item {[17]}
A. Mostafazadeh, J. Math. Phys. {\bf 35}, 1095 (1994).
\item {[18]}
A. Mostafazadeh, Int. J. Mod. Phys. {\bf A 11}, 2941 (1996).
\item {[19]}
G. Barnich, M. Henneaux and C. Schomblond, Phys. Rev.
{\bf D 44}, R939 (1991).
\item {[20]}
G. Esposito, G. Gionti and C. Stornaiolo, Nuovo Cimento
{\bf B 110}, 1137 (1995).
\item {[21]}
H. Ozaki, HEP-TH 0010273.
\item {[22]}
I. V. Kanatchikov, Int. J. Theor. Phys. {\bf 40}, 1121 (2001).
\item {[23]}
I. V. Kanatchikov, GR-QC 0012038.
\item {[24]}
C. Rovelli, GR-QC 0111037;
C. Rovelli, GR-QC 0202079; C. Rovelli, GR-QC 0207043.

\bye